\documentclass[prl%
reprint, 
twocolumn,
secnumarabic,
tightenlines,
amssymb,
amsmath,
nobibnotes,
nofootinbib, 
floatfix,
aps, 
showpacs,showkeys,
]{revtex4-2}
\usepackage{hyperref}
\usepackage{multirow}%
\expandafter \ifx \csname package@font\endcsname\relax\else
\expandafter \expandafter \expandafter
\usepackage
\expandafter \expandafter
\expandafter{\csname package@font\endcsname}%
\fi
\usepackage{natbib} 
\usepackage{ulem} 

\usepackage{textcomp}
\usepackage{gensymb}

\usepackage{forest}
\usepackage{orcidlink}

\usepackage{hyperref} %
\newcommand{\etal}{\textit{et al.}}

\usepackage{cleveref}
\usepackage{graphicx} %

\begin{document}

\title{Re-optimization of a deep neural network model for electron–carbon scattering using new experimental data}

\author{Beata E. Kowal\,\orcidlink{0000-0003-3646-1653}}
\email{beata.kowal@uwr.edu.pl}

\author{Krzysztof M. Graczyk\,\orcidlink{0000-0002-0038-6340}}
\email{krzysztof.graczyk@uwr.edu.pl}

\author{\\Artur M. Ankowski\,\orcidlink{0000-0003-4073-8686}}
\author{Rwik Dharmapal Banerjee\,\orcidlink{0000-0003-3639-7532}}
\author{Jose L. Bonilla\,\orcidlink{0009-0009-3240-1494}}
\author{Hemant Prasad\,\orcidlink{0009-0003-1897-9616}}
\author{Jan T. Sobczyk\,\orcidlink{0000-0003-4991-2919}}

\affiliation{Institute of Theoretical Physics, University of Wroc\l aw, plac Maxa Borna 9,
50-204, Wroc\l aw, Poland}

\date{\today}%

\begin{abstract}
We present an updated deep neural network model for inclusive electron–carbon scattering. Using the bootstrap model [Phys.Rev.C 110 (2024) 2, 025501] as a prior, we incorporate recent experimental data, as well as older measurements in the deep inelastic scattering region, to derive a re-optimized posterior model. We examine the impact of these new inputs on model predictions and associated uncertainties. Finally, we evaluate the resulting cross-section predictions in the kinematic range relevant to the Hyper-Kamiokande and DUNE experiments.
\end{abstract}

\maketitle

\section{Introduction}

Accurate modeling of nuclear effects in (anti)neutrino-nucleus scattering is crucial for studies of fundamental properties of neutrinos~\cite{doi:10.1146/annurev-nucl-102115-044720,NuSTEC:2017hzk}. Indeed, their description is currently one of the primary sources of systematic uncertainty in measurements of (anti)neutrino oscillation parameters, including the charge-parity symmetry violating phase in the lepton sector~\cite{coyle2025neutrinonucleuscrosssectionimpacts}.

Neutrinos and electrons interact with atomic nuclei in ways that exhibit significant similarities. Interpreting (anti)neutrino scattering data is challenging due to flux averaging and the different interaction mechanisms that contribute to the same final states. Electron-scattering data, however, are much easier to understand; being collected for a fixed beam energy, they are pretty informative even at the inclusive level~\cite{Ankowski:2022thw}.

There is a broad consensus that transferring knowledge from electron to neutrino scattering physics is both feasible and beneficial. Such knowledge transfer is expected to significantly reduce theoretical uncertainties in (anti)neutrino–nucleus cross-section calculations~\cite{Ankowski:2022thw}. This reduction is crucial for next-generation neutrino oscillation experiments, which aim to probe neutrino properties with unprecedented precision~\cite{DUNE:2020lwj,Hyper-KamiokandeProto-:2015xww}.

A significant effort has been devoted to better understanding how (anti)neutrinos interact with atomic nuclei. The typical approach involves formulating a theoretical framework and comparing its predictions with experimental measurements,  which usually leads to modifications of the description and/or adjustments of its parameters. 

We adopt a different perspective. Specifically, we aim to develop a fully data-driven model for predicting nuclear cross sections using artificial intelligence (AI) methods. Our work begins with a study of electron–nucleus scattering~\cite{Kowal:2023dcq,Graczyk:2024pjm}. Similar approaches are discussed in the papers by Al~Hammal~\etal~\cite{AlHammal:2023svo} and Sobczyk~\etal~\cite{Sobczyk:2024ajv}, where neural networks are employed to model electron–nucleus cross sections. 

Neural network-based techniques are increasingly being used across various domains of physics~\cite{MEHTA20191,Graczyk:2020hih,GraczykMatyka2023}, including particle and nuclear physics~\cite{Radovic:2018dip}. In the case of (anti)neutrino–nucleus interactions, deep learning techniques are applied to generate neutrino-nucleus scattering events~\cite{Bonilla:2025wir,ElBaz:2023ijr,ElBaz:2025qjp}, model neutrino-nucleus cross sections \cite{hackett2025machinelearningneutrinonucleuscross}, unfolding of neutrino measurements~\cite{sp1f-n9k2} as well as to extract the axial form factor of the nucleon in a model-independent way~\cite{Alvarez-Ruso:2018rdx}.

In this paper, we continue the development of the deep neural network (DNN) model for inclusive electron–carbon scattering cross sections, initially introduced in Ref.~\cite{Kowal:2023dcq}. The available electron–carbon scattering data span a broad kinematic range, offering valuable insights into nuclear effects. The carbon target, which is structurally similar to oxygen, is also relevant to the Hyper-Kamiokande experiment~\cite{Hyper-KamiokandeProto-:2015xww}.

Notably, we have demonstrated that the DNN model trained on inclusive electron–carbon scattering data can be effectively extended—using transfer learning techniques—to describe electron scattering off other nuclear targets, including oxygen, aluminum, calcium, iron, lithium, and even light nuclei such as helium-3~\cite{Graczyk:2024pjm}.

We anticipate that a similar transfer learning approach could be applied to develop a model for electron–argon scattering, based on the electron–carbon model. This would be particularly valuable, as argon is the target material used in the DUNE experiment~\cite{DUNE:2020jqi}.

Our approach is entirely data-driven, in contrast to those explored in the papers~\cite{Christy:2007ve,Bosted:2007xd,Bodek:2022gli,Bodek:2023dsr,Bodek:2024mrp}, where longitudinal and transverse components of the electron-nucleus cross sections are extracted from the data using some theoretical constraints. Another example of an empirical-based approach is the superscaling approach \cite{Donnelly:1999sw,Megias:2016lke,Gonzalez-Jimenez:2014eqa,Gonzalez-Rosa:2024udj}. We intentionally omit any theoretical assumptions to obtain fully model-independent predictions of nuclear cross sections.

Interest in the study of electron-nucleus interactions has been growing within the neutrino interaction community in recent years. Monte Carlo (MC) generator models, originally developed for (anti)neutrino-nucleus interactions, have been adapted to test theoretical models of neutrino interactions against electron-nucleus scattering data. An example of such an initiative is the e-GENIE project~\cite{electronsforneutrinos:2020tbf}, together with the CLASS and $e4\nu$ collaboration~\cite{CLAS:2021neh}. Other collaborations, such as eWro~\cite{Zmuda:2015rea}, Neut~\cite{Abe:2024avs}, GIBUU~\cite{Mosel:2023zek}, and Achilles \cite{Isaacson:2025cnk}, have also made significant efforts by developing MC electron-nucleus models and comparing their predictions with electron scattering data.

A key advantage of deep learning models is their ability to efficiently update with new data through re-optimization procedures. In this work, we incorporate recent measurements from the Mainz experiment~\cite{Mihovilovic:2024ymj}, as well as higher-energy data from Gomez~\etal~\cite{Gomez:1993ri}. As a result, we produce uncertainty maps that, for a given incident electron energy, quantify the model’s predictive uncertainty for inclusive cross sections across the allowed kinematic region. These maps are generated for energies relevant to neutrino experiments such as Hyper-Kamiokande and DUNE. 

The paper is organized as follows: Section~\ref{Sec:framework} introduces the method briefly, while Section~\ref{Sec:results} discusses the obtained results. Our conclusions are included in Section~\ref{Sec:conclusions}. In the Appendix~\ref{Appendix_1}, we give the optimization parameter settings.

\section{Framework}
\label{Sec:framework}

\begin{figure}
\begin{center}
\includegraphics[width=0.5\textwidth]{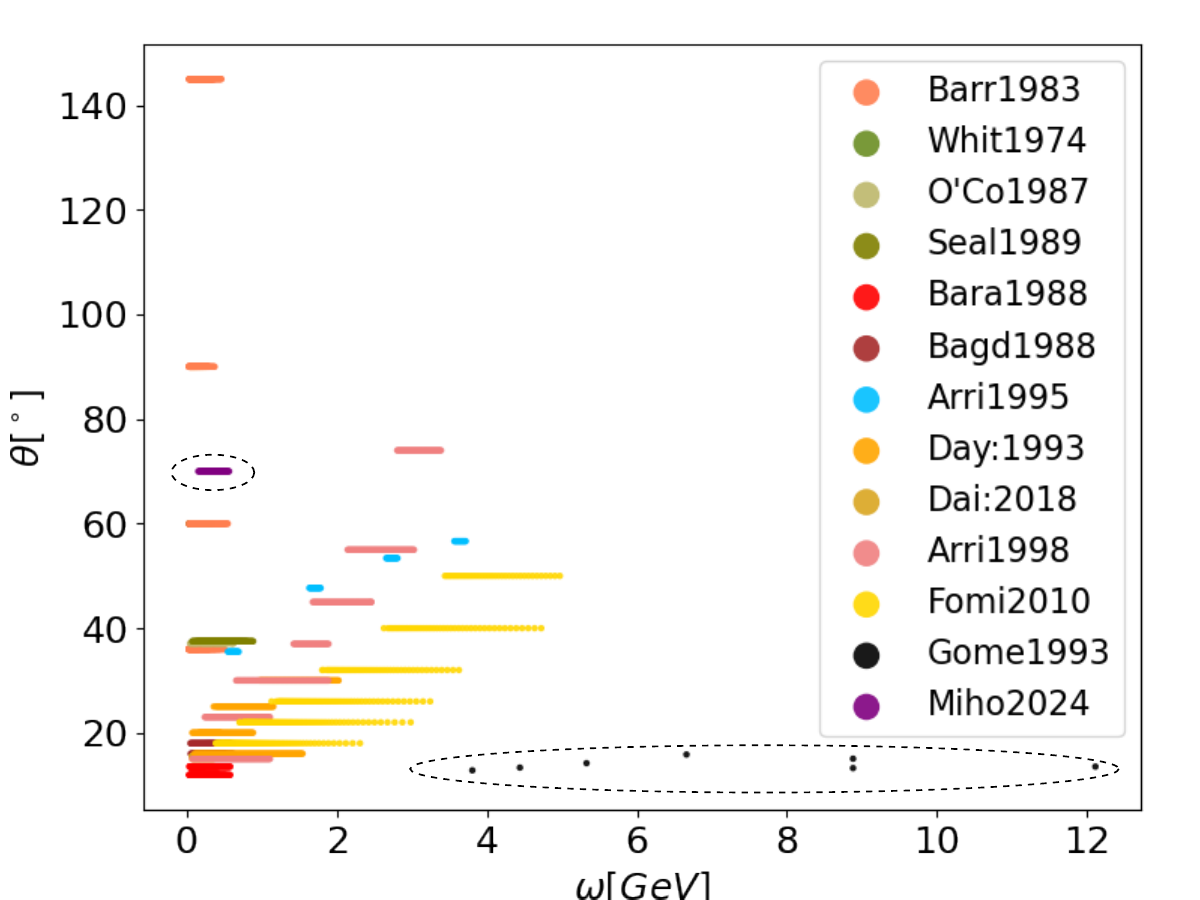}
\includegraphics[width=0.5\textwidth]{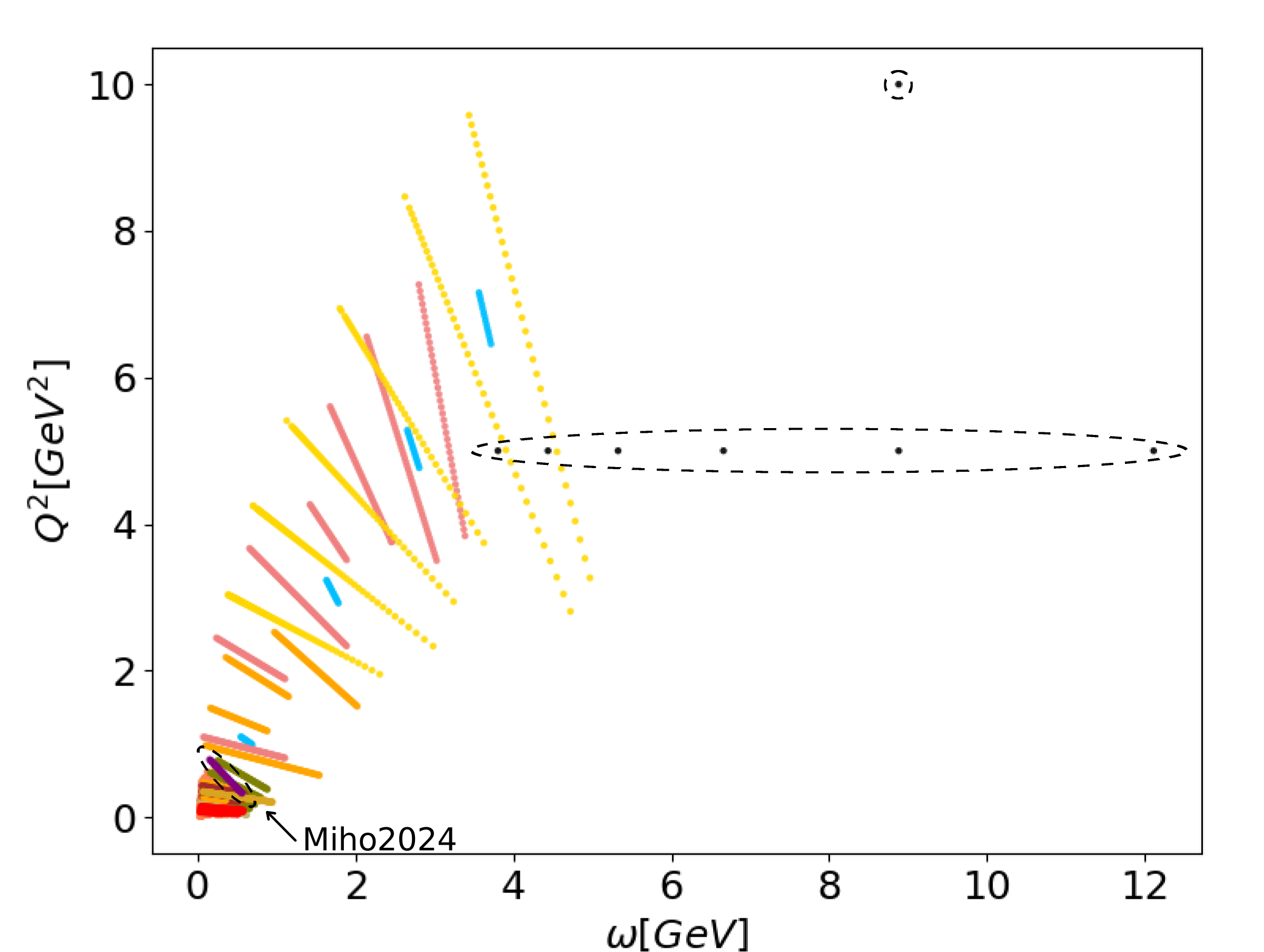}
\end{center}
\caption{Kinematic domain covered by the previous and present analyses. The top and bottom panels show the data in the $(\omega,\theta)$ and $(\omega, Q^2)$ variables, respectively.    New data used to update the cross-section model are enclosed within two ellipses. By $\theta$ and $\omega$, the scattering angle and transfer of energy are denoted,
$Q^2=2E (E-\omega) (1-\cos\theta)$, where $E$ is the electron beam energy.} 
\label{fig:data_domain}
\end{figure}

\subsection{Data}

Our idea is to use the previously developed model for inclusive electron–carbon scattering cross sections~\cite{Kowal:2023dcq} as prior information. We refine this model using new measurements~\cite{Mihovilovic:2024ymj} and additional data that were not included in the previous analysis~\cite{Gomez:1993ri}. The original model was constructed using eleven independent datasets listed in Table~\ref{tab:syserrors}. As described in Ref.\cite{Kowal:2023dcq}, low energy transfer data points are excluded from the analysis by applying a kinematic cut~\cite{Ankowski:2014yfa}.
\begin{table}[!tb]
\caption{The data utilized in the analysis. The numbers of points refer to the data surviving the kinematic cut. The first two datasets were not included in the previous analysis~\cite{Kowal:2023dcq}. }
\label{tab:syserrors}
\begin{ruledtabular}
\begin{tabular}[t]{llrr}
\multirow{2}{*}{Reference} & \multirow{2}{*}{Abbrev.}  &  Norm. & Number \\
& & uncert. & of points\\
 \hline
Mihovilovic~\etal~\cite{Mihovilovic:2024ymj} & Miho2024 &  $2\%$   & 54 \\ 
Gomez~\etal~\cite{Gomez:1993ri} &   Gome1993 &    $0.6\%$ &   7 \\
 \hline
 \hline
Arrington \etal~\cite{Arrington:1995hs} & Arri1995 &  $4.0\%$  & 56 \\
Arrington \etal~\cite{Arrington:1998ps} & Arri1998 &  $4.0\% $  &  398 \\
Bagdasaryan \etal~\cite{Bagdasaryan:1988hp} & Bagd1988 & $ 10.0\%$  & 125 \\
Baran \etal~\cite{Baran:1988tw} & Bara1988 & $ 3.7 \%$  &  259 \\
Barreau \etal~\cite{Barreau:1983ht} & Barr1983 & $ 2.0\%$  &  1243 \\
Dai \etal~\cite{Dai:2018xhi} & Dai2018 & $ 2.2 \%$  &  177 \\
Day \etal~\cite{Day:1993md} & Day1993 & $3.4\%$  &  316 \\
Fomin \etal~\cite{Fomin:2010ei} & Fomi2010  & $ 4.0\%$  &  359 \\ 
O’Connell \etal~\cite{O'Connell:1987ag} & O’Con1987 & $ 5.0\%$  & 51 \\
Sealock \etal~\cite{ Sealock:1989nx} & Seal1989 & $ 2.5\%$  & 250 \\
Whitney \etal~\cite{Whitney:1974hr} & Whit1974 & $ 3.0\%$  & 31 \\
\hline
Total &  &  & 3265 + 61
\end{tabular}
\end{ruledtabular}
\end{table}

In Fig.~\ref{fig:data_domain}, we present the kinematic domain covered by the measurements considered in this study, along with the data included in earlier analyses. As shown, the data from Ref.~\cite{Mihovilovic:2024ymj} span the quasi-elastic and more inelastic regions, at a four-momentum transfer squared of about $Q^2=0.8$~GeV$^2$, the scattering angle of $70^\circ$, and energy transfer ranging from $0$ to $0.6$~GeV. These measurements provide new insights into the nuclear cross sections. In this region,  a few prior data sets are available, such as those from Barreau~\etal~\cite{Barreau:1983ht} and Sealock~\etal~\cite{Sealock:1989nx}. In contrast, the measurements by Gomez~\etal~\cite{Gomez:1993ri} are located in the deep inelastic scattering (DIS) region, where no other data currently exist. These points, therefore, serve as the primary source of information on the cross sections in this kinematic range.

\subsection{DNN model}

Our model consists of an ensemble of fifty neural networks. Each network consists of ten blocks, each containing $300$ hidden units, and a batch normalization layer. For more detailed information, see Section III.3.2. of Ref.~\cite{Kowal:2023dcq}.

Each neural network in the ensemble predicts the normalized differential cross sections for specific electron energy ($E$), energy transfer ($\omega$), and scattering angle ($\theta$) values. To train a network, the measurements were rescaled by the factor: $(10^9 /(E\cos\frac{\theta}{2})) \sigma_{Mott}$, where
\begin{equation}
  \sigma_{Mott}   =  \frac{\displaystyle \alpha^2 \cos^2\frac{\theta}{2} }{\displaystyle 4 E^2 \sin^4\frac{\theta}{2}}. 
    \label{scaling}
\end{equation}


\subsection{Statistical approach}
To find the optimal parameters of the neural networks, we search for the minimum of the loss function
\begin{equation}
\label{Eq:chi2_tot}
\chi_\text{tot}^2 = \sum_{k=1}^{N_{tot}} \left[ \chi_k^2(\lambda_k) + \frac{1}{2}\left(\frac{1 - \lambda_k }{\Delta \lambda_k}\right)^2\right],
\end{equation}
which consists of $\chi^2_k s$ for every independent data set, namely,
\begin{equation}
\label{Eq:chi2}
\chi_k^2(\lambda_k) = \frac{1}{2} \sum_{i=1}^{N_k} 
\left(\frac{d \sigma_{k}^i -  \lambda_k d \sigma_k^\text{i,net}}{\Delta d \sigma_{k}^i}\right)^2,
\end{equation}
where $\Delta d \sigma_{k}^i$ is the statistical and uncorrelated systematic uncertainty for the $i$-th point in the $k$-th dataset, $d \sigma_{k}^i$ and $\sigma_k^\text{i,net}$ are measured and predicted by model cross sections for $i$-th point in $k$-th dataset, respectively. In the previous analysis, we considered $N_{tot}=11$ independent datasets, while in the present study, we have $N_{tot}=13$. For each dataset, we distinguish the overall systematic normalization uncertainty $\Delta \lambda_k$ and a corresponding normalization parameter $\lambda_k$. The information regarding the data can be found in Table~\ref{Table:chi2_and_lambdas}. To optimize the values of $\lambda_k$, we implement the algorithm proposed in Refs. \cite{Graczyk:2011kh,Graczyk:2014lba,Graczyk:2014coa}. Tables~\ref{Table:chi2_and_lambdas} and \ref{Table:chi2_and_lambdas_test} collect the values of $\chi^2_{nor} \equiv \chi^2$ [per number of points in the given dataset] obtained for the bootstrap model.

In our previous paper, we considered two statistical approaches: Monte Carlo dropout and the bootstrap model. The latter demonstrated superior extrapolation capabilities. Therefore, in the present work, we adopt the bootstrap approach~\cite{Tibshirani_96,Breiman1996}, which is an example of an ensemble method~\cite{survey_uncertainty_in_dnn}, allowing us to easily estimate predictive uncertainties and prevent overfitting.

As mentioned above, our model consists of $50$ DNNs, each trained on a distinct bootstrap replica of the experimental data. The bootstrap method, rooted in frequentist statistics~\cite{Efron_bootstrap}, enables the generation of these replicas through appropriate sampling. Despite its frequentist foundation, it often yields predictions comparable to those obtained via Bayesian approaches~\cite{10.1214/12-AOAS571}.

The model's prediction is obtained by averaging the outputs of all ensemble members, while the square root of the variance across these outputs provides an estimate of the predictive uncertainty. This approach captures both the uncertainty stemming from the experimental measurements and the variability introduced by the model parameters. Moreover, because predictions are made through an ensemble, the overall result remains robust even in cases where individual models may overfit the data.

In practice, to generate the bootstrap dataset, each data point with central value \( d\sigma_k^i \) and uncertainty \( \Delta d\sigma_k^i \) is used to produce a new bootstrap sample:
\[
d\sigma_{k,\text{bootstrap}}^i = d\sigma_k^i + r \, \Delta d\sigma_k^i,
\]
where \( r \) is drawn from a standard normal distribution.

In the previous analysis, we split the whole dataset into training and test subsets in a 9:1 ratio. For the present analysis, we use the same split—the training and test data points from the previous study are preserved in the corresponding datasets of this work. Additionally, we split the data from Ref. \cite{Mihovilovic:2024ymj} into training and test sets using the same 9:1 ratio. For the dataset from Ref.~\cite{Gomez:1993ri}, which contains only a few data points located out of the rest of the measurements, we include all of them in the training dataset\footnote{While we initially employed a 9:1 split for the data from Ref.~\cite{Gomez:1993ri}, the resulting performance in this kinematic region was found to be suboptimal.}.

\begin{table}
\caption{ The prior and posterior values of normalization parameter $\lambda_{prior}$ and $\lambda_{posterior}$, see Eq.~\ref{Eq:chi2_tot} as well as $\chi^2_{nor}$ (divided by number of data points) for prior and posterior analyses computed for training datasets. \label{Table:chi2_and_lambdas} }
\begin{ruledtabular}
\begin{tabular}{ l|c |c|r|r} 
 \hline
 \textbf{Dataset} &   $\lambda_{prior}$ & $\lambda_{posterior}$ & $\chi^2_{nor,prior}$ & $\chi^2_{nor,posterior}$ \\ 
  \hline 
   \hline
Miho2024 &   -  & 1.0004 &  17.711   & 1.799  \\   
 \hline
Gome1993 &  -  & 0.9929  & 393.384  & 9.364 \\ 
 \hline
 \hline 
 Arri1995 &   1.0096 & 0.9847 & 0.279 & 0.456\\ 
  \hline
Arri1998  & 0.9998 & 0.9723 & 0.250 & 0.898\\ 
  \hline
Bagd1988 &  1.0273 &  1.0239 &  0.151 & 0.174\\ 
  \hline
Bara1988  &   1.0090 &  1.0070 & 0.178 & 0.254\\ 
  \hline
Barr1983 &   0.9889 & 0.9963 & 0.311 & 0.842\\
  \hline
Dai2018   &   1.0002 & 1.0011 & 0.117 & 0.457\\  
  \hline
Day1993   &   0.9885 &  0.9884 & 0.350 & 0.784\\ 
  \hline
Fomi2010  &   1.0083 & 0.9786 & 0.208 & 0.818\\ 
  \hline
O’Con1987  &   1.0249 & 1.0100 & 0.370 & 0.378\\ 
  \hline
Seal1989   &  1.0176 & 1.0033 & 0.267 & 0.300\\ 
  \hline
Whit1974   &  0.9282 & 0.9626 & 7.441 & 7.584\\   
 \hline
\end{tabular}
\end{ruledtabular}
\end{table}
\begin{table}
\caption{\raggedright{The $\chi^2_{nor}$ values (divided by number of data points) for prior and posterior analyses computed for the test dataset. \label{Table:chi2_and_lambdas_test} }}
\begin{ruledtabular}
\begin{tabular}{ l|r|r} 
 \hline
 \textbf{Dataset}  & $\chi^2_{nor,prior}$ & $\chi^2_{nor,posterior}$ \\ 
  \hline 
   \hline
Miho2024  & 7.956 & 0.399 \\   
 \hline
Gome1993  &     -  & -    \\ 
 \hline
 \hline 
Arri1995  &  0.172   & 0.379  \\ 
  \hline
Arri1998  &  0.415  &  1.276 \\ 
  \hline
Bagd1988  & 0.123 & 0.105\\  
  \hline
Bara1988   &  0.176  & 0.229 \\  
  \hline
Barr1983  & 0.403   &  1.018 \\ 
  \hline
Dai2018   &  0.267 & 0.327\\  
  \hline
Day1993  & 0.483  & 0.946 \\ 
  \hline
Fomi2010  &  0.201  & 0.729 \\ 
  \hline
O’Con1987 & 0.246 & 0.093 \\  
  \hline
Seal1989  &  0.335  & 0.272 \\ 
  \hline
Whit1974  & 3.112   & 2.613 \\    
 \hline
\end{tabular}
\end{ruledtabular}
\end{table}

\section{Results}
\label{Sec:results}

Each DNN in the ensemble was individually re-optimized, starting from the prior configuration of the weights (i.e., neural network parameters). Parameters (weights) in all layers were updated to minimize the total loss defined in Eq.~\ref{Eq:chi2_tot}. Unlike the prior optimization, which ran the optimizer for approximately 60,000 epochs, the present analysis limits the number of epochs to less than 1,000. The details of the study are given in the Appendix~\ref{Appendix_1}.

\begin{figure}
\begin{center}
\includegraphics[width=0.5\textwidth]{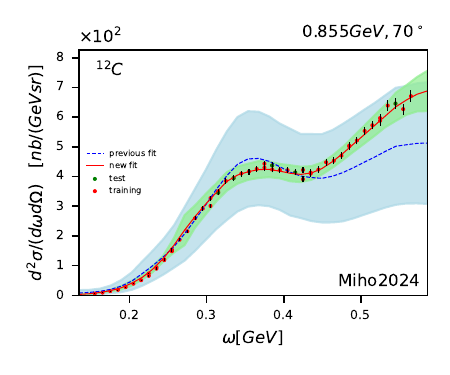}
\end{center}
\caption{Double-differential cross section \(\frac{d^2 \sigma}{d\omega\, d\Omega}\) for inclusive electron scattering on carbon. The red line represents the posterior model predictions, with the associated \(1\sigma\) uncertainty shown as a green shaded region. The blue dashed line denotes the prior model predictions~\cite{Kowal:2023dcq}, and the light blue area indicates their \(1\sigma\) uncertainty. The red points correspond to the training dataset, while the green points indicate the test dataset. The data from Mihovilovic~\etal~\cite{Mihovilovic:2024ymj}. In the top left corner, we specify the incoming electron energy $E$  and scattering angle $\theta$.} 
\label{fig:Miho2024}
\end{figure}

The inclusion of two additional datasets significantly enhances the constraints on the nuclear cross sections in the kinematic region where these data were collected. As shown in Fig.~\ref{fig:Miho2024}, we compare the measurements from Ref.~\cite{Mihovilovic:2024ymj} with both the previous and the current predictions of our models. While the earlier fit was already in reasonable agreement with the new data, the updated model presented in this work shows a better agreement and a substantially reduced predictive uncertainty.

The Mihovilovic \textit{et al.} data fall within the region already covered by older measurements, but are concentrated in a relatively narrow part of phase space. Because these points are characterized by tiny statistical and normalization uncertainties, the fit has limited flexibility in this restricted domain, which results in a comparatively larger $\chi^2_{nor}$ for the training dataset ($\sim 1.8$). In contrast, the $\chi^2_{nor}$ for the corresponding test dataset is lower ($\sim 0.4$) and consistent with the other datasets (excluding Gomez \textit{et. al}), indicating that the model does not overfit the Mainz points and performs reliably in prediction. Increasing the effective systematic uncertainty would reduce the training–test discrepancy, while broader kinematic coverage in future Mainz measurements would improve the global constraints.

The agreement between our previous model and the measurements by Gomez~\etal~\cite{Gomez:1993ri} is within two standard deviations (see Fig.~\ref{fig:Gome1993}). Including these measurements in the analysis improves the agreement to within one standard deviation. Nevertheless, additional measurements in this region are needed to further constrain the model. It is also worth noting that, for some of the data points, the predictive uncertainties remain comparable to those in the previous fit.

The tables \ref{Table:chi2_and_lambdas} and \ref{Table:chi2_and_lambdas_test} provide a quantitative summary of the complete analysis. They contain the normalization parameters ($\lambda_i$'s) as well as the $\chi^2$ values for each dataset for the previous and present studies. The worst metrics are obtained for the data from Whitney~\etal~\cite{Whitney:1974hr}, as shown in Fig.~\ref{Fig:Barr1983:3}. In this case, neither model can accurately capture the low-energy transfer data (with tiny uncertainties).

We also verified that the train--test comparison shows consistent performance across datasets, confirming that the model generalizes well and does not overfit the training data.

Note that we report $\chi^{2}_{nor}$ per point as a qualitative diagnostic of the model performance. 
A formal interpretation of $\chi^{2}_{nor} \approx 1$ is not expected here, because the usual assumptions for a $\chi^{2}$ goodness-of-fit test are not satisfied: the network has a large number of effective parameters compared to the number of data points  and each dataset includes an independent normalization nuisance that reduces the effective tension. Therefore, $\chi^{2}_{nor}$ should be understood as a comparative indicator across datasets rather than as a strict statistical test.

{At the same time, extremely small $\chi^{2}_{nor}$ values could indicate possible overfitting. In our training samples, the lowest $\chi^{2}_{nor}$ is obtained for the  Bagdasaryan~\etal~\cite{Bagdasaryan:1988hp} dataset (see Figs.~\ref{Fig:5} and \ref{Fig:6}), 
which can be attributed to the relatively large quoted uncertainties for all its measurements, allowing the network to achieve a low residual error. Among the test datasets, the smallest $\chi^{2}_{nor}$ corresponds to O’Connell \etal~\cite{O'Connell:1987ag} dataset (see also Fig.~\ref{Fig:Barr1983:4}). 
This dataset contains only a single series of measurements (about five points) with small uncertainties which follow the general trend established by the training dataset and other experiments.  Consequently, its low $\chi^{2}_{nor}$ value is consistent with good predictive behavior rather than overfitting.

Finally, we observe that for the previously used data, the new model typically yields similar predictions as the previous one, as shown in Figs.~\ref{Fig:Barr1983:1}-\ref{Fig:7}. We notice minor differences between the previous and present models for the Barreau~\etal~\cite{Barreau:1983ht} data, as shown in Figs.~\ref{Fig:Barr1983:1} and \ref{Fig:Barr1983:2}. However, both models' predictions are in agreement at a one-sigma level.

The primary motivation for our study stems from neutrino physics. Specifically, we aim to assess how well the electron–nucleus cross sections are understood within the kinematic range relevant to neutrino oscillation experiments. With a nuclear cross section model at hand, we are now able to investigate this question quantitatively.
\begin{figure*}
  \includegraphics[width=0.8\textwidth]{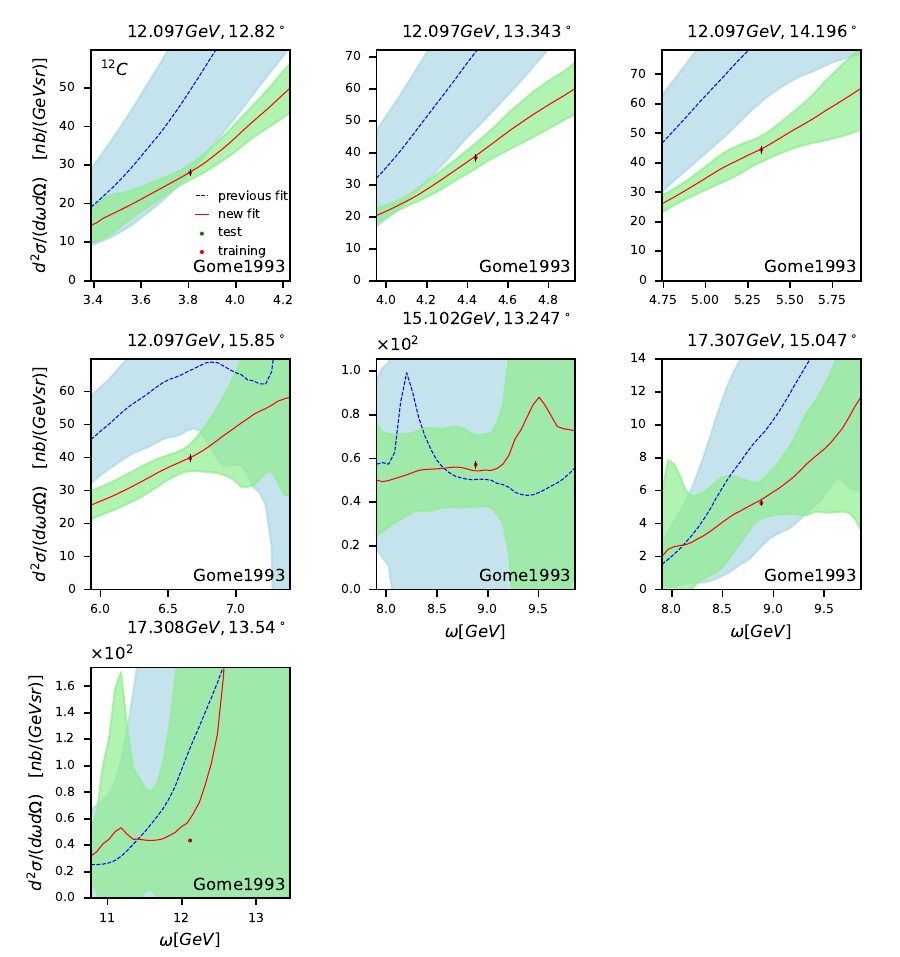}
  \caption{Same as in Fig.~\ref{fig:Miho2024} but for the data from: Gomez~\etal~\cite{Gomez:1993ri}.}
  \label{fig:Gome1993}
\end{figure*} 

In Figs.~\ref{fig_map_0.6} and \ref{fig_map_2.5}, we present maps of predictive uncertainties for the inclusive electron-carbon cross-section model, calculated for energies of \(E=0.6\) GeV and \(2.5\) GeV, respectively. These energies correspond to the peak neutrino energies of the Hyper-Kamiokande and DUNE experiments. The displayed uncertainties are accompanied by contour lines that illustrate the kinematic domains covered by both experiments.  Contours enclose 68\% of the events obtained from a  MC simulation of charged current neutrino-carbon (Hyper-Kamiokande) and neutrino-argon (DUNE) scattering using the NuWro generator~\cite{Golan:2012wx}.

\begin{figure*} 
\includegraphics[width=0.8\textwidth]{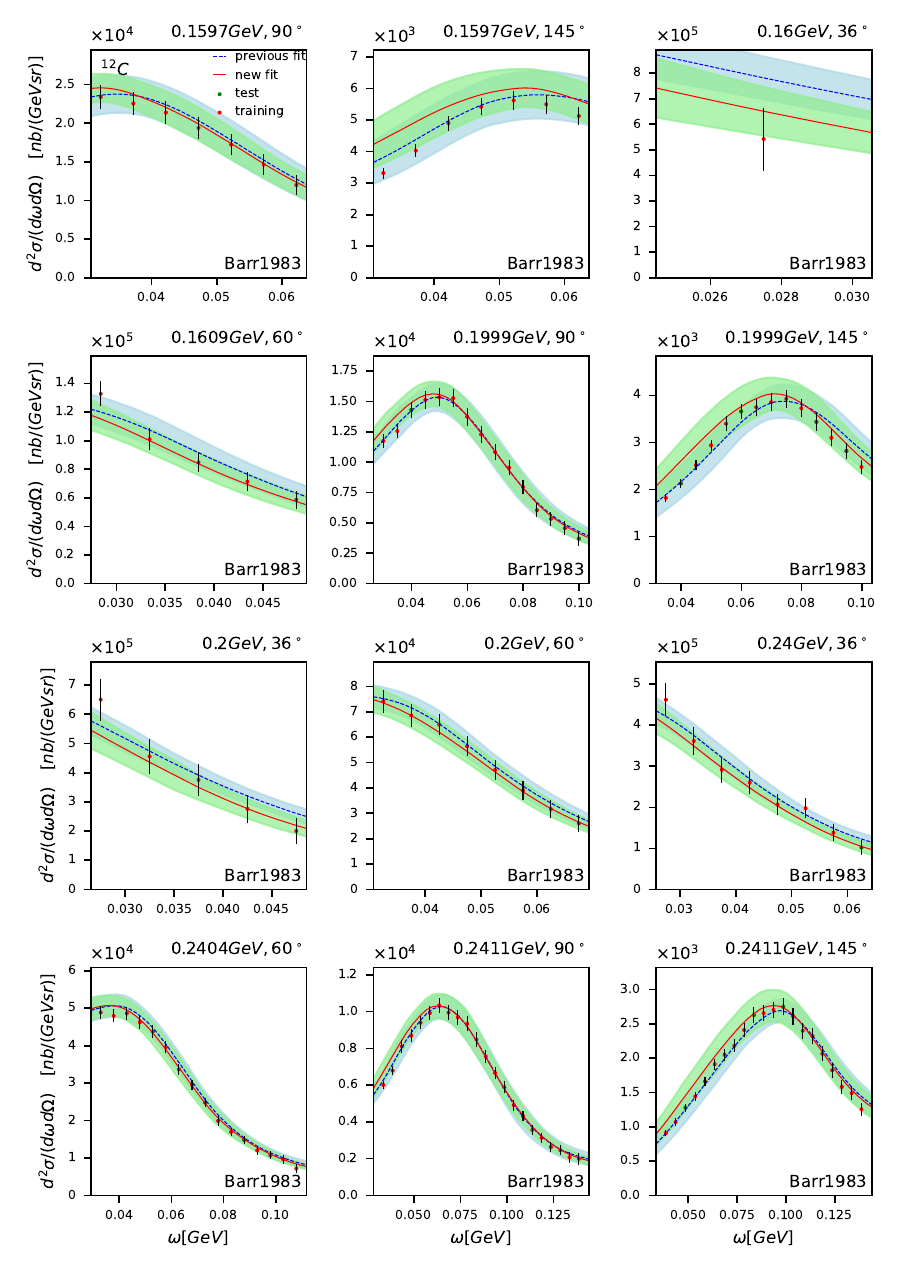}
\caption{Same as in Fig.~\ref{fig:Miho2024} but for the data from: Barreau~\etal~\cite{Barreau:1983ht} [Barr1983].}
\label{Fig:Barr1983:1}
\end{figure*}
\begin{figure*}
  \includegraphics[width=0.8\textwidth]{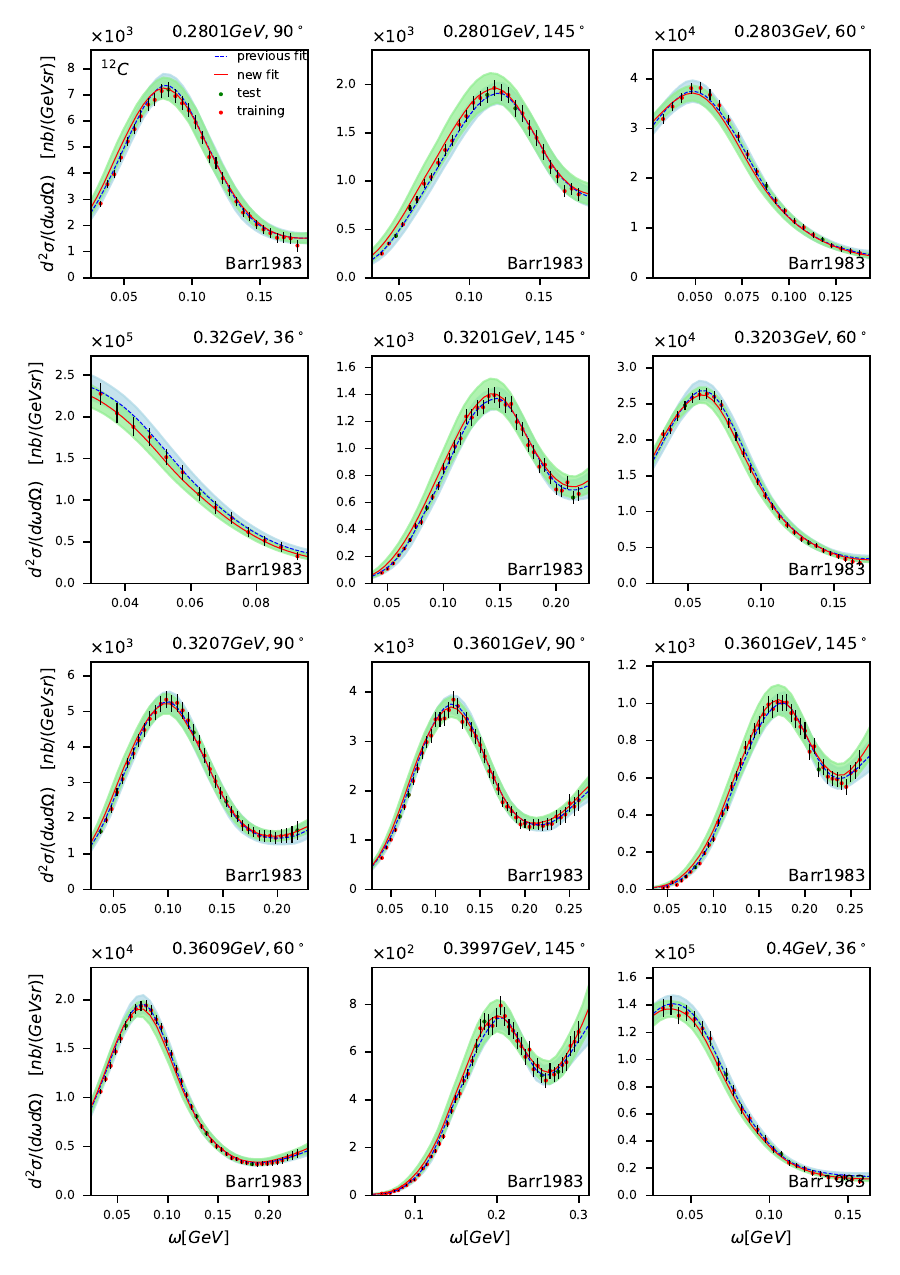}
  \caption{Same as in Fig.~\ref{fig:Miho2024} but for the data from: Barreau~\etal~\cite{Barreau:1983ht} [Barr1983].}
\label{Fig:Barr1983:2}
\end{figure*}

Neutrino interactions in Hyper-Kamiokande will be dominated by quasielastic scattering and $\Delta(1232)$ resonance excitation. In this regime, our model exhibits relatively low uncertainties for scattering angles between $30^\circ$ and $100^\circ$ across all energy transfers. The lowest\footnote{The lowest uncertainty, in the contour, is about 7.8\%  for $\theta=35^\circ$.} uncertainties—below $20$\%—are observed in the angular ranges of $30^\circ$–$40^\circ$, $60^\circ$–$70^\circ$, and $85^\circ$–$90^\circ$. As shown by the contour plots, the electron scattering cross sections are known to within approximately $20$\%  across most of the kinematic range. However, at larger scattering angles, above $100^\circ$, the uncertainties in the electron scattering cross sections exceed $100$\%.

The DUNE experiment will operate at higher neutrino energies, allowing it to probe a broader range of kinematic regimes—including quasielastic, resonance, and more inelastic interactions—compared to the Hyper-Kamiokande experiment. The DNN model used in this analysis demonstrates the lowest predictive uncertainties\footnote{The lowest uncertainty, in the contour, is about 6.5\%  for $\theta=35^\circ$.}, of the order of $10$\% for energy transfers between approximately 
$0.5$ and $1.5$~GeV and for scattering angles between  $15^\circ$ and  $30^\circ$. However, certain kinematic configurations in the DUNE data are associated with higher uncertainties in the DNN predictions, particularly for lower energy transfers ($0-1.5$~GeV) and larger scattering angles, ranging from $ 30^\circ$ to $40^\circ$.

\section{Conclusions}
\label{Sec:conclusions}

The deep neural network model for inclusive electron–carbon cross sections has been re-optimized to include new measurements. In this process, the previously developed model~\cite{Kowal:2023dcq} was incorporated as prior information. The updated model can be extended to other nuclear targets using transfer learning techniques~\cite{Graczyk:2024pjm}. Notably, the model is entirely data-driven and does not rely on any theoretical assumptions. This approach allows us to assess the precision of nuclear cross-section predictions in the kinematic region relevant to the Hyper-Kamiokande and DUNE experiments. Currently available electron-scattering data can constrain models of nuclear effects used in neutrino experiments at the 10–20\% level. However, since Hyper-Kamiokande and DUNE require neutrino cross sections to be known at the few-percent level, our findings highlight an urgent need for systematic electron-scattering studies across the relevant kinematic range. In particular, measurements at high scattering angles ($\theta > 100^\circ$) and energy transfers around $250$~MeV are especially important for improving nuclear cross-section knowledge in the context of Hyper-Kamiokande. For DUNE, measurements at low energy transfer and low scattering angles are most critical.

The DNN model is available from the GitHub repository~\cite{neuwro}.

\appendix

\section{Details of optimization}
\label{Appendix_1}

Training of each neural network model in the ensemble did not exceed 1,000 epochs. Training was performed using a minibatch configuration with 2 to 4 batches. We used the AdamW optimization algorithm ($\beta_1=0.9$, $\beta_2=0.999$, $\epsilon=10^{-7}$, weight decay $= 0.004$) with a starting learning rate $lr_0=0.0005$. The base learning rate $lr_0$ was reduced by a factor of $5$ every $200$ epochs.

\section{Model prediction vs. data}

In this section, in Figs.~\ref{Fig:Barr1983:1}, \ref{Fig:Barr1983:2}, \ref{Fig:Barr1983:3}, \ref{Fig:Barr1983:4}, \ref{Fig:5}, \ref{Fig:6}, and \ref{Fig:7}, we present a comparison of our DNN model predictions with measurements taken into account in our previous paper~\cite{Kowal:2023dcq}.  

\begin{figure*}
  \includegraphics[width=0.8\textwidth]{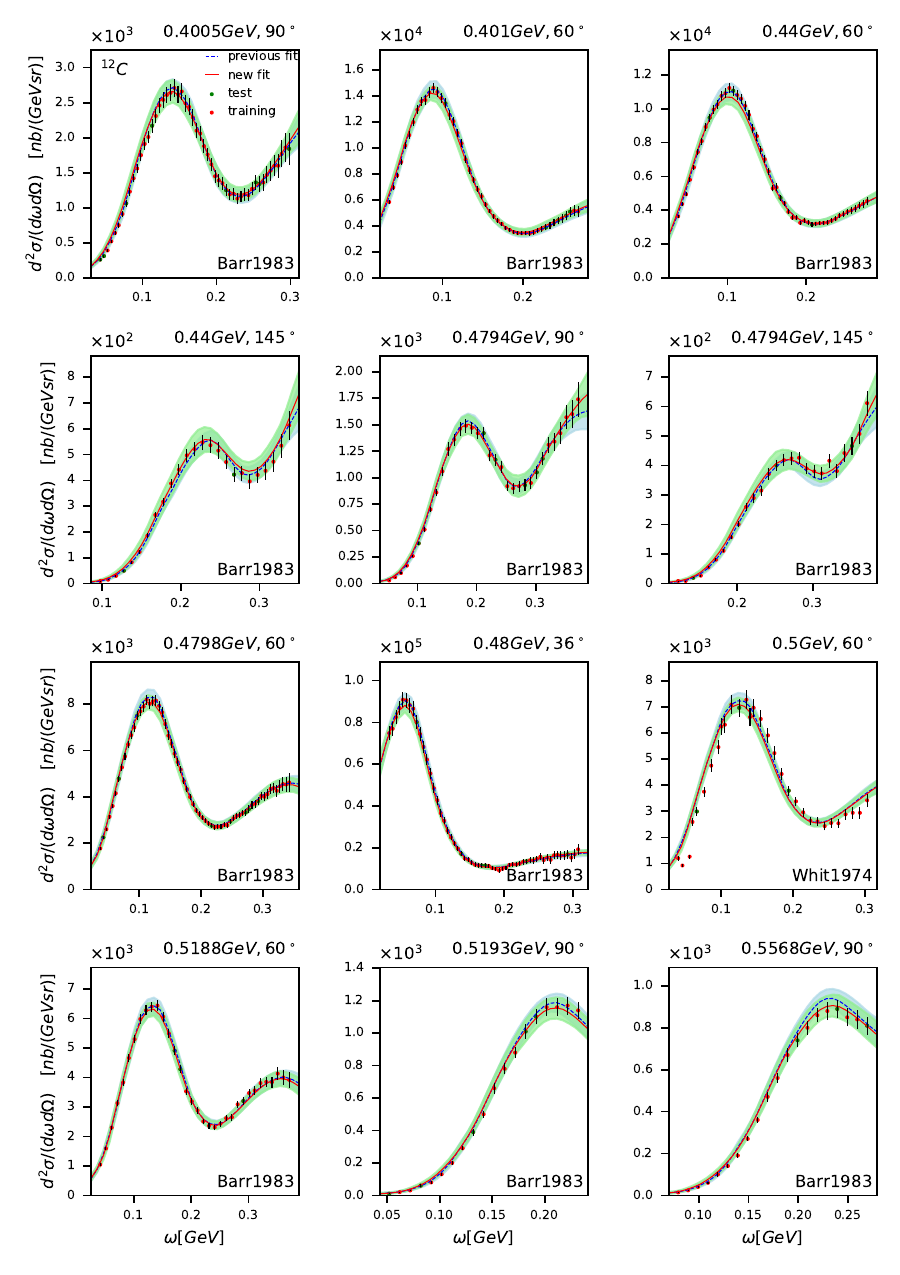}
  \caption{Same as in Fig.~\ref{fig:Miho2024} but for the data from: Barreau~\etal~\cite{Barreau:1983ht} [Barr1983] and Whitney~\etal~\cite{Whitney:1974hr} [Whit1974].}
\label{Fig:Barr1983:3}
\end{figure*}
\begin{figure*}
  \includegraphics[width=0.8\textwidth]{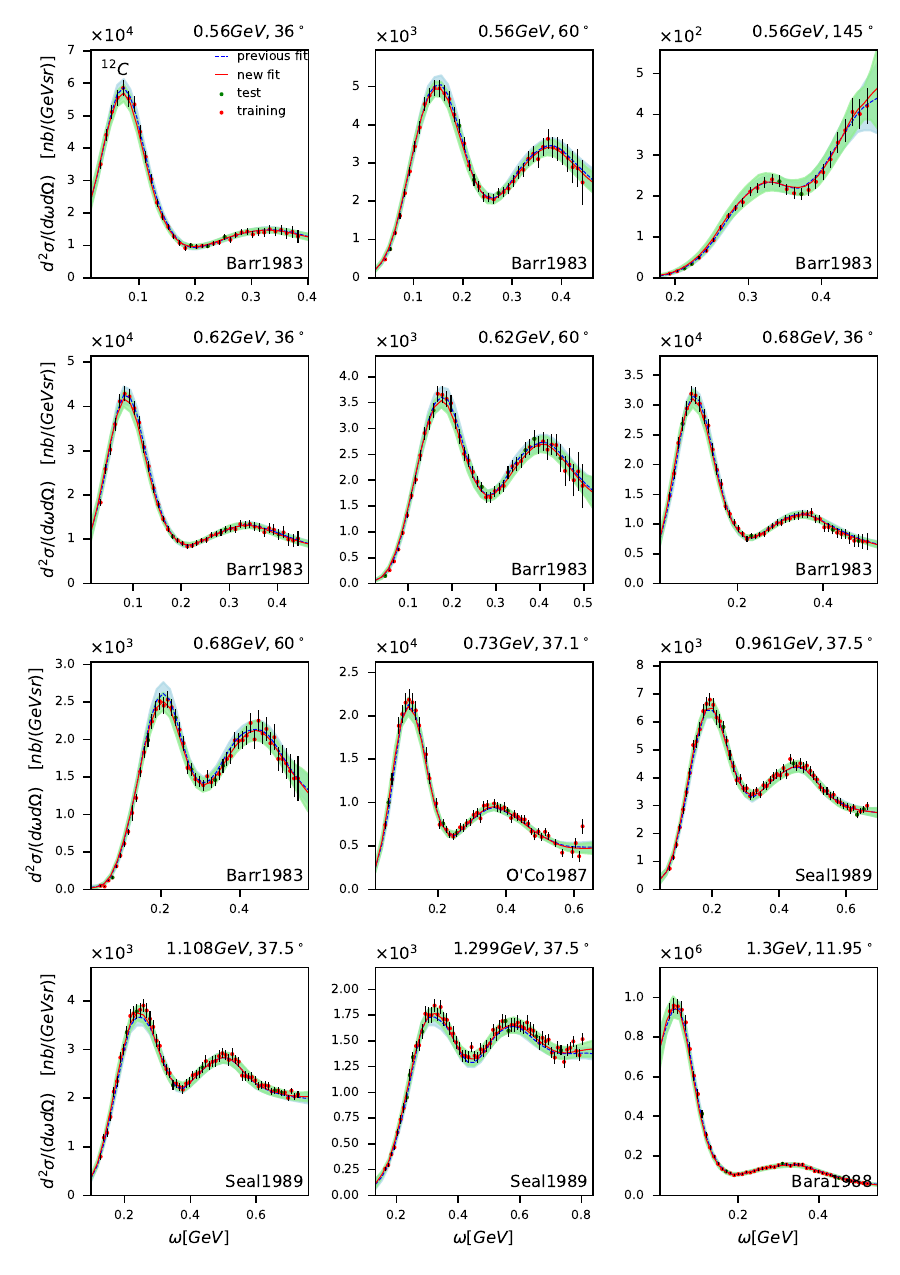}
  \caption{Same as in Fig.~\ref{fig:Miho2024} but for the data from: Barreau~\etal~\cite{Barreau:1983ht} [Barr1983], 
  Baran~\etal~\cite{Baran:1988tw} [Bara1988], O'Connell~\etal~\cite{O'Connell:1987ag} [O’Con1987] and Sealock~\etal~\cite{ Sealock:1989nx} [Seal1989].}
  \label{Fig:4}
\label{Fig:Barr1983:4}
\end{figure*}
\begin{figure*}
  \includegraphics[width=0.8\textwidth]{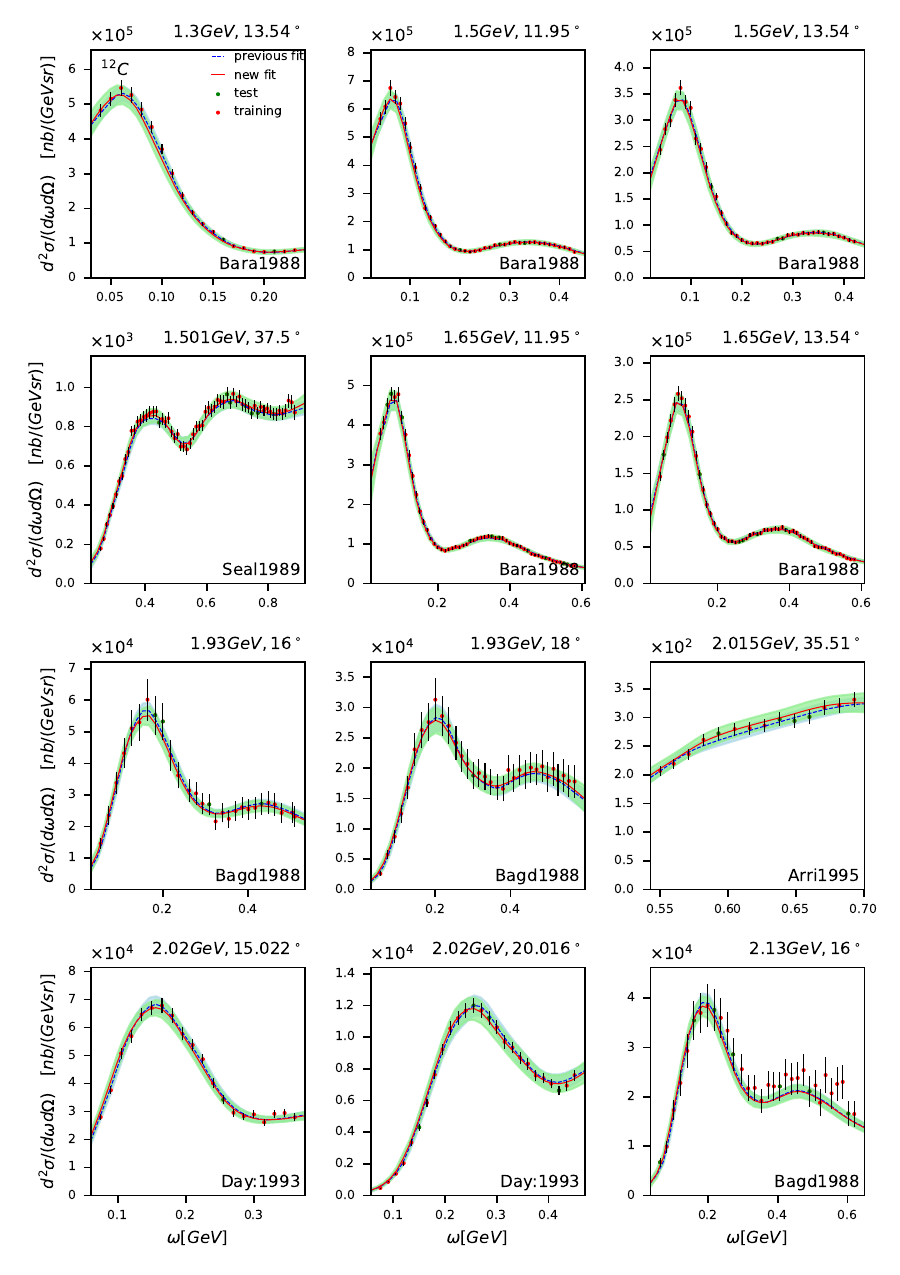}
  \caption{Same as in Fig.~\ref{fig:Miho2024} but for the data from: Baran~\etal~\cite{Baran:1988tw} [Bara1988], Bagdasaryan~\etal~\cite{Bagdasaryan:1988hp} [Bagd1988] and Sealock~\etal~\cite{ Sealock:1989nx} [Seal1989] and Arrington~\etal~\cite{Arrington:1995hs} [Arri1995], and  Day~\etal~\cite{Day:1993md} [Day1993].}
  \label{Fig:5}
\end{figure*}
\begin{figure*}
  \includegraphics[width=0.8\textwidth]{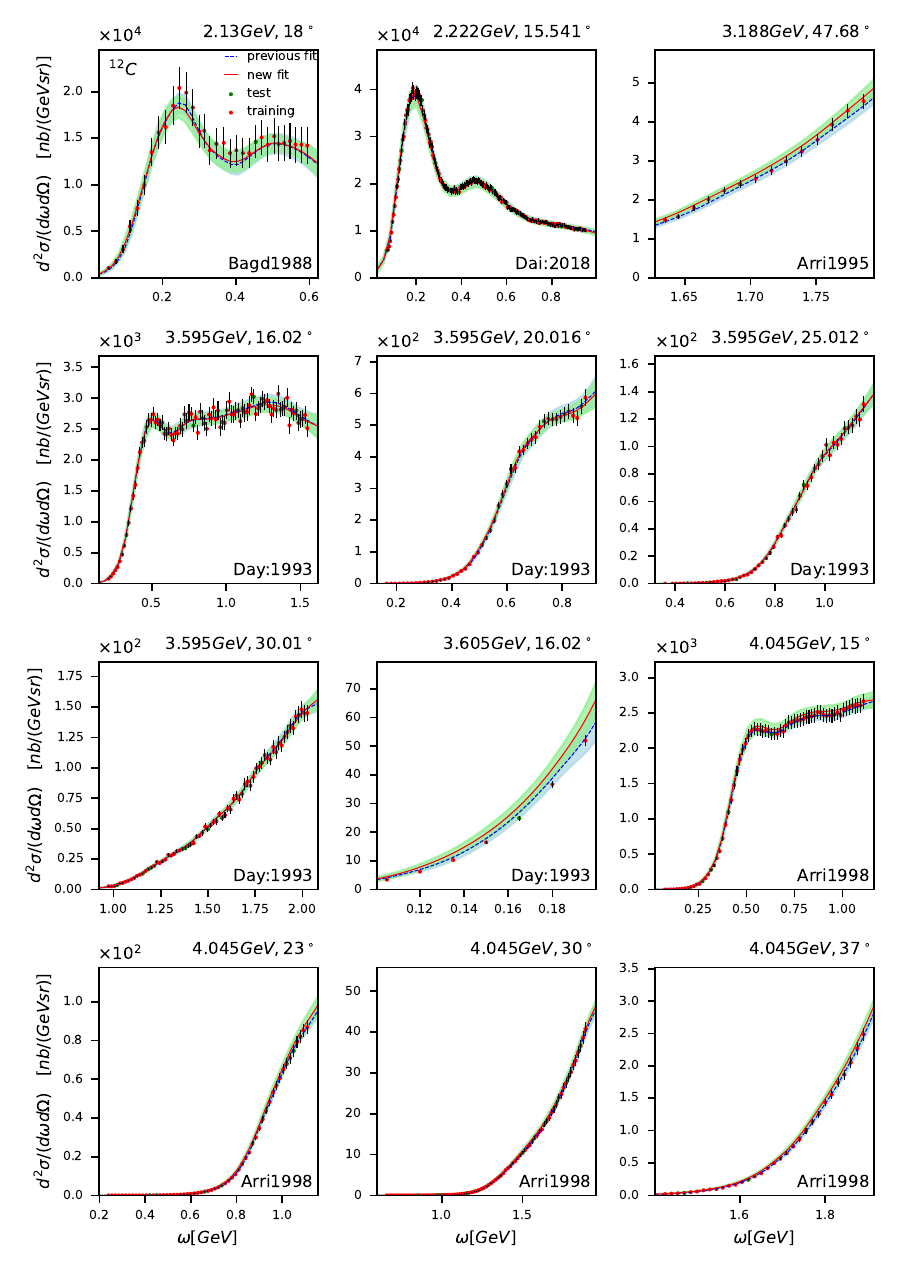}
  \caption{Same as in Fig.~\ref{fig:Miho2024} but for the data from: Bagdasaryan~\etal~\cite{Bagdasaryan:1988hp} [Bagd1988], Dai~\etal~\cite{Dai:2018xhi} [Dai2018], Arrington~\etal~\cite{Arrington:1995hs} [Arri1995], Arrington~\etal~\cite{Arrington:1998ps} [Arri1998], and Day~\etal~\cite{Day:1993md} [Day1993].}
  \label{Fig:6}
\end{figure*}
\begin{figure*}
  \includegraphics[width=0.8\textwidth]{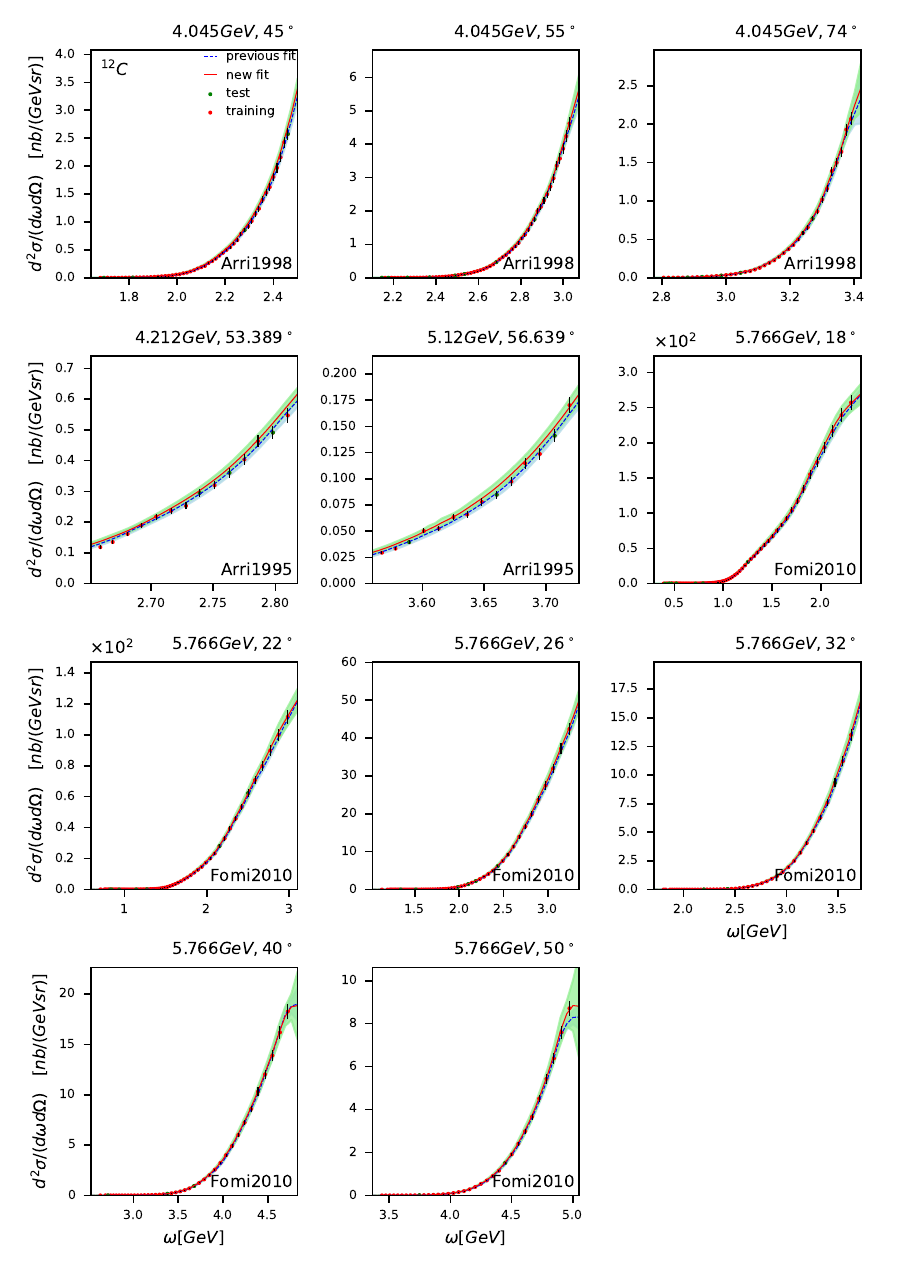}
  \caption{Same as in Fig.~\ref{fig:Miho2024} but for the data from: Arrington~\etal~\cite{Arrington:1995hs} [Arri1995], Arrington~\etal~\cite{Arrington:1998ps} [Arri1998], and Fomin~\etal~\cite{Fomin:2010ei} [Fomi2010].}
  \label{Fig:7}
\end{figure*}

\begin{figure}
    \includegraphics[width=0.5\textwidth]{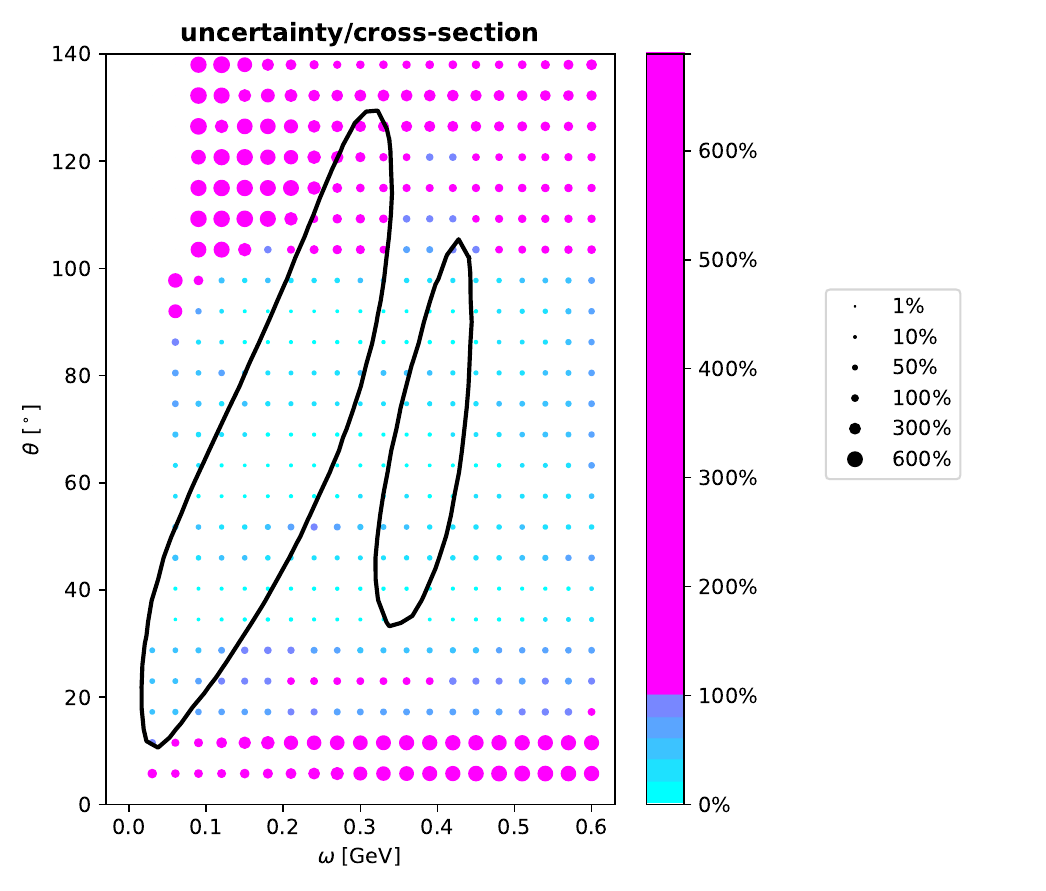}\\
  \caption{Maps of predictive uncertainties for the DNN model for electron energy $E=0.6$~GeV corresponding to the peak energy of Hyper-Kamiokande. The contours enclose 68\% of charged-current neutrino-carbon scattering events generated by NuWro.}
  \label{fig_map_0.6}
\end{figure}
\begin{figure}
    \includegraphics[width=0.5\textwidth]{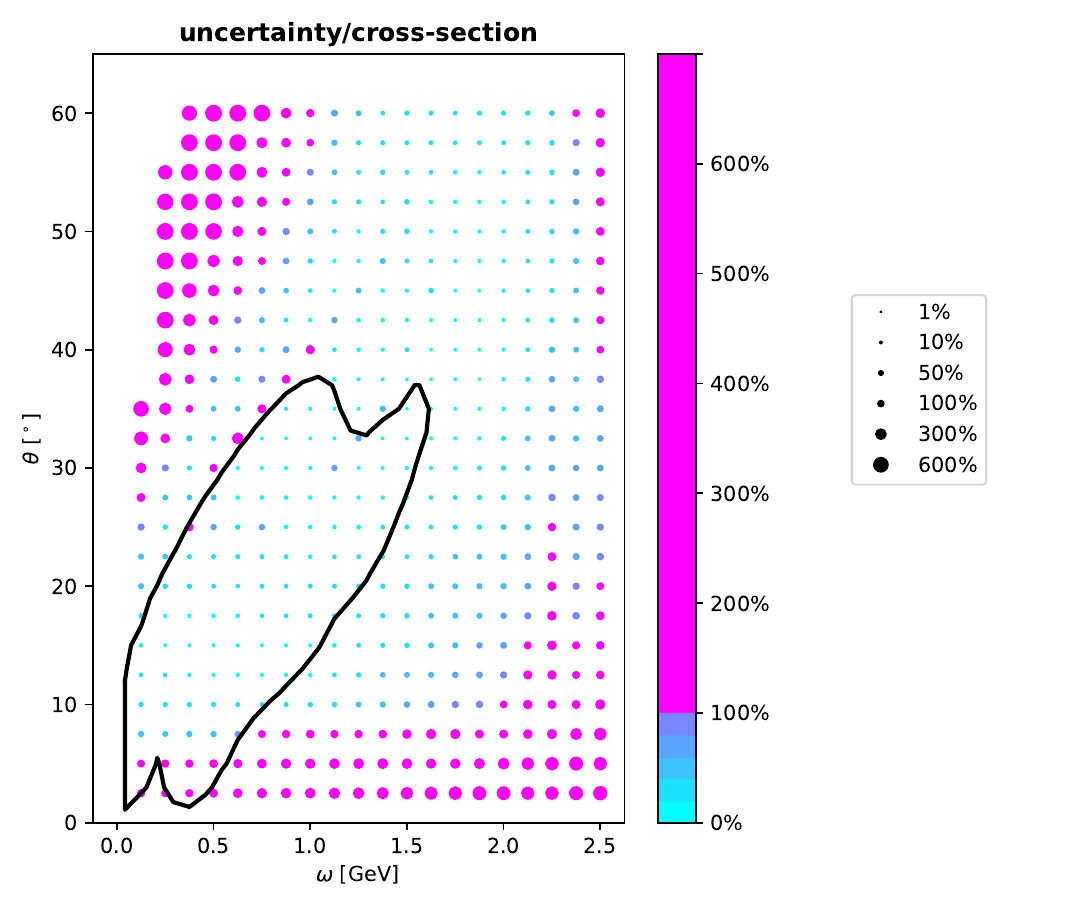}
  \caption{Maps of predictive uncertainties for the DNN model for electron energy $E=2.5$~GeV corresponding to the peak energy of Dune. The contour encloses 68\% of charged-current neutrino-argon scattering events generated by NuWro.}
  \label{fig_map_2.5}
\end{figure}

\newpage

\begin{acknowledgments}
This research is partly (K.M.G., A.M.A., J.T.S.) or fully (B.E.K., J.L.B, H.P., R.D.B.) supported by the Na{\-}tional Science Centre under grant UMO-2021/41/B/ST2/ 02778. K.M.G is partly supported by the ``Excellence Initiative – Research University" for the years 2020-2026 at the University of Wroc\l aw.
\end{acknowledgments}

\normalem
\bibliographystyle{apsrev4-2}
\bibliography{bibdata,bibdrat,kgm_papers}

\end{document}